\documentclass[useAMS,usenatbib]{mn2e}
\usepackage{txfonts}
\usepackage{natbib}
\usepackage{graphicx}
\usepackage{multirow}
\usepackage{ulem}

\newcommand{\apj}{ApJ}
\newcommand{\apjl}{ApJ Let.}
\newcommand{\aj}{AJ}

\newcommand{\aap}{A\&A}

\newcommand{\mnras}{MNRAS}

\newcommand{\araa}{ARA\&A}

\newcommand{\iaucirc}{IAU~Circ.}	
\newcommand{\na}{New Astron.}
\newcommand{\kms}{km~s$^{-1}$}

\newcommand{\oiii}{\textrm{[O\,{\sc iii}]}}

\newcommand{\nai}{\textrm{Na\,{\sc i}}}

\newcommand{\nii}{\textrm{N\,{\sc ii}}}
\newcommand{\niif}{\textrm{[N\,{\sc ii}]}}
\newcommand{\msun}{{\rm\ M}_\odot}

\title[The Morphology of the Expanding Ejecta of V2491 Cygni (2008 N.2)]{The Morphology of the Expanding Ejecta of V2491 Cygni (2008 N.2)}
\author[V. A. R. M. Ribeiro et~al.]
       {V. A. R. M. Ribeiro,$^{1}$\thanks{E-mail: var@astro.livjm.ac.uk}
         M. J. Darnley,$^{1}$
         M. F. Bode,$^{1}$
         U. Munari,$^{2,3}$
         \newauthor
         D. J. Harman,$^{1}$
         I. A. Steele$^{1}$
         and J. Meaburn$^{4}$ \\
         $^{1}$ Astrophysics Research Institute, Liverpool John Moores University, Twelve Quays House, Egerton Wharf, Birkenhead, CH41 1LD \\
         $^{2}$ INAF Astronomical Observatory of Padova, via dell'Osservatorio, 36012 Asiago (VI), Italy \\
         $^{3}$ ANS Collaboration, c/o Astronomical Observatory, 36012 Asiago (VI), Italy \\
         $^{4}$ Jodrell Bank Centre for Astrophysics, University of Manchester, Oxford Road, Manchester, M13 9PL}

\begin{document}

\date{Accepted 2010 November 8.  Received 2010 November 8; in original form 2010 September 22}

\pagerange{\pageref{firstpage}--\pageref{lastpage}} \pubyear{2002}

\maketitle

\label{firstpage}

\begin{abstract}
Determining the evolution of the ejecta morphology of novae provides valuable information on the shaping mechanisms in operation at early stages of the nova outburst. Understanding such mechanisms has implications for studies of shaping for example in proto-Planetary Nebulae. Here we perform morpho-kinematical studies of V2491 Cyg using spectral data to determine the likely structure of the ejecta and its relationship to the central system and shaping mechanisms. We use {\sc shape} to model different morphologies and retrieve their spectra. These synthetic spectra are compared with observed spectra to determine the most likely morphology giving rise to them, including system inclination and expansion velocity of the nova ejecta. We find the best fit remnant morphology to be that of polar blobs and an equatorial ring with an implied inclination of 80$^{+3}_{-12}$ degrees and an maximum expansion velocity of the polar blobs of 3100$^{+200}_{-100}$ \kms\ and for the equatorial ring 2700$^{+200}_{-100}$ \kms. This inclination would suggest that we should observe eclipses  which will enable us to determine more precisely important parameters of the central binary. We also note that the amplitude of the outburst is more akin to the found in recurrent nova systems.
\end{abstract}


\section{Introduction}\label{intro}
A classical nova (CN) outburst occurs in an interacting binary system comprising a main-sequence or evolved star (the secondary), which fills its Roche Lobe, and a white dwarf (WD; the primary). Matter transferred from the secondary to the WD builds up on the surface. Once extensive CNO cycle nuclear burning commences under these degenerate conditions, a thermonuclear runaway occurs ejecting $\sim$ 10$^{-5}-$10$^{-4}$ $\msun$ of matter from the WD surface with velocities from hundreds to thousands of \kms\ \citep[see, e.g.][and references therein]{BE08,B10}. A related sub-class of CNe are the recurrent novae (RNe), which unlike CNe, recur on time-scales of order tens of years. Such a short recurrence time-scale is usually attributed to a high mass WD, probably close to the Chandrasekhar limit, together with a high accretion rate \citep{SST85,YPS05}.

Nova Cygni 2008 N.2 (hereafter, V2491 Cyg) was discovered on 2008 April 10.8 UT (taken as $t$ = 0) at about 7.7 mag on unfiltered CCD frames \citep{NBJ08} and its nature confirmed spectrometrically yielding velocities of $\sim$ 4500 \kms\ \citep[H$\alpha$ FWHM,][]{AM08}. \citet{HM08} identified USNO-B1.0 1223-042965 as the likely progenitor system (within 0.9 arc seconds of V2491 Cyg). An archival search by \citet{JM08} of the Asiago Schmidt telescope images of the progenitor collected over the period 1970-1986 showed no previous outburst. \citet{MSD10} have provided a detailed study of the photometric and spectrometric evolution of V2491 Cyg, including a photo-ionization analysis of the ejecta and their chemical abundances. They derived the reddening for V2491 Cyg of $E(B-V)$ = 0.23$\pm$0.01, a distance of 14 kpc, and $V$ = 17.9, $R_c$ = 17.5, $I_c$ = 17.1 for the progenitor in quiescence. 

\begin{figure}
\resizebox{\hsize}{!}{\includegraphics{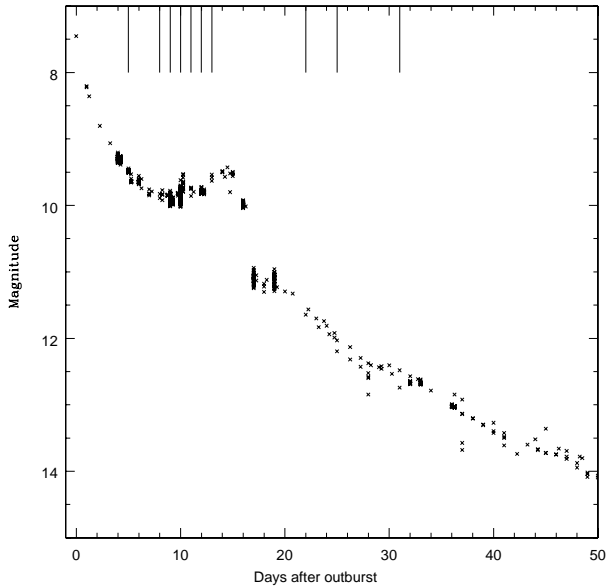}}
\caption{Optical light curve for V2491 Cyg from AAVSO data. The vertical lines are days where data were taken with the Meaburn Spectrograph on the 2-meter robotic Liverpool Telescope.}
\label{fig:light_curve}
\end{figure}

In Fig. \ref{fig:light_curve} we show the light curve for the first 50 days after outburst from the AAVSO\footnote{See http://www.aavso.org/}. A rapid decline is observed for the first 10 days after outburst and then a sudden rebrightening peaking around day 15. It is not fully understood why this rebrightening should happen. For example, \citet{HK09} have modelled the rebrightening by introducing magnetic activity as an additional energy source to nuclear burning. Although, a magnetic WD has a spin closely related to the binary orbital period \citep{KWF90}, taken previously by several authors as 0.0958 days \citep{BPB08}, but there is no observed evidence of a short periodicity in the X-rays related to the orbital period \citep{IKO09,POE10} and generally polars are weak X-ray sources \citep{KW87}. It is also noteworthy that A. Baklanov (private communication) is not confident that the above period is related to the orbital period of the system, a matter that is addressed in \citet{DRB10}.

V2491 Cyg is of particular interest because it was detected as an X-ray source pre-outburst \citep{IK08,IKB08,IKO09}. This is only the second CN to be detected in X-rays pre-outburst \citep[after V2487 Ophiuchi,][]{HS02}. V2487 Oph was originally suggested to be a RN because of its rapid decline in the optical domain and the presence of a plateau phase during decline \citep[see e.g.,][]{HKK02,HS02}. V2487 Oph's true recurrent nature came to light only after a search of the Harvard College Observatory archival photographic collection revealed an outburst on 1900 June 20 \citep{PSX09}. They used two methods to determine a recurrence timescale of order $\sim$ 20 years, a Monte Carlo simulation to calculate the probability that a given recurrence timescale would produce exactly two discovered eruptions and by directly estimating the most likely number of eruptions. Furthermore, V2491 Cyg has very similar proprieties to the suggested extragalactic RN M31N 2007-12b, which was shown to have similar magnitude and  colour to the RN RS Oph at quiescence \citep{BDS09}. On the other hand, \citet{IKO09} derived 0.2$-$10 keV X-ray luminosities ranging from 10$^{34}-$4$\times$10$^{35}$ erg s$^{-1}$ implying inter-outburst mass accretion rates in the range of 10$^{-9}$ to 10$^{-8}$ $\msun$ yr$^{-1}$ for a 1 $\msun$ WD. Such low values for the accretion rate would imply a recurrence timescale of $\gtrsim$ 100 yrs \citep{YPS05,IKO09,POE10}. However, a higher WD mass will reduce the recurrence time.

Post-outburst {\it Swift} observations of V2491 Cyg showed a clearly detected X-ray source, although the count rate was almost an order of magnitude fainter than during pre-outburst observations \citep{POE10}. The post-outburst observations showed three distinct phases: first a hard X-ray emission phase, then a super-soft source (SSS) phase and finally the decline of the SSS with the hard component again becoming prominent, since it fades much less rapidly than the SSS emission \citep{POE10}.

For decades it has been known that spectral line profiles contain a great deal of information about the expanding remnant of a nova outburst \citep[e.g.,][]{H72}. Here we use spectrometric data to constrain the structure and velocity field for the expanding remnant of V2491 Cyg. In Section \ref{obs}, we describe our observations and data reduction methodology. The results are presented in Section \ref{results} and then modelled in Section \ref{modelling}. In Section \ref{results2} we present the results of the modelling and in Section \ref{discussion} we discuss the implications of the results.

\section{Observations and Data Reduction}\label{obs}
Optical spectra of V2491 Cyg were obtained with the Meaburn Spectrograph (a thermocooled, Apogee AP7p 512$\times$512 pixel thinned Tektronix array prototype fibre-fed robotic spectrograph, now superseded by FRODOSpec) on the 2-metre robotic Liverpool Telescope \citep[LT; ][]{SSR04} sited at the Observatorio del Roque de Los Muchachos on the Canary Island of La Palma, Spain. The spectrograph was fed using a close-packed fibre bundle input array, consisting of 49 $\times$ 0.8$^{\prime\prime}$  diameter fibres which were reformatted as a slit with the fibres in random order.

\begin{figure*}
  \includegraphics[angle=270, width=170mm]{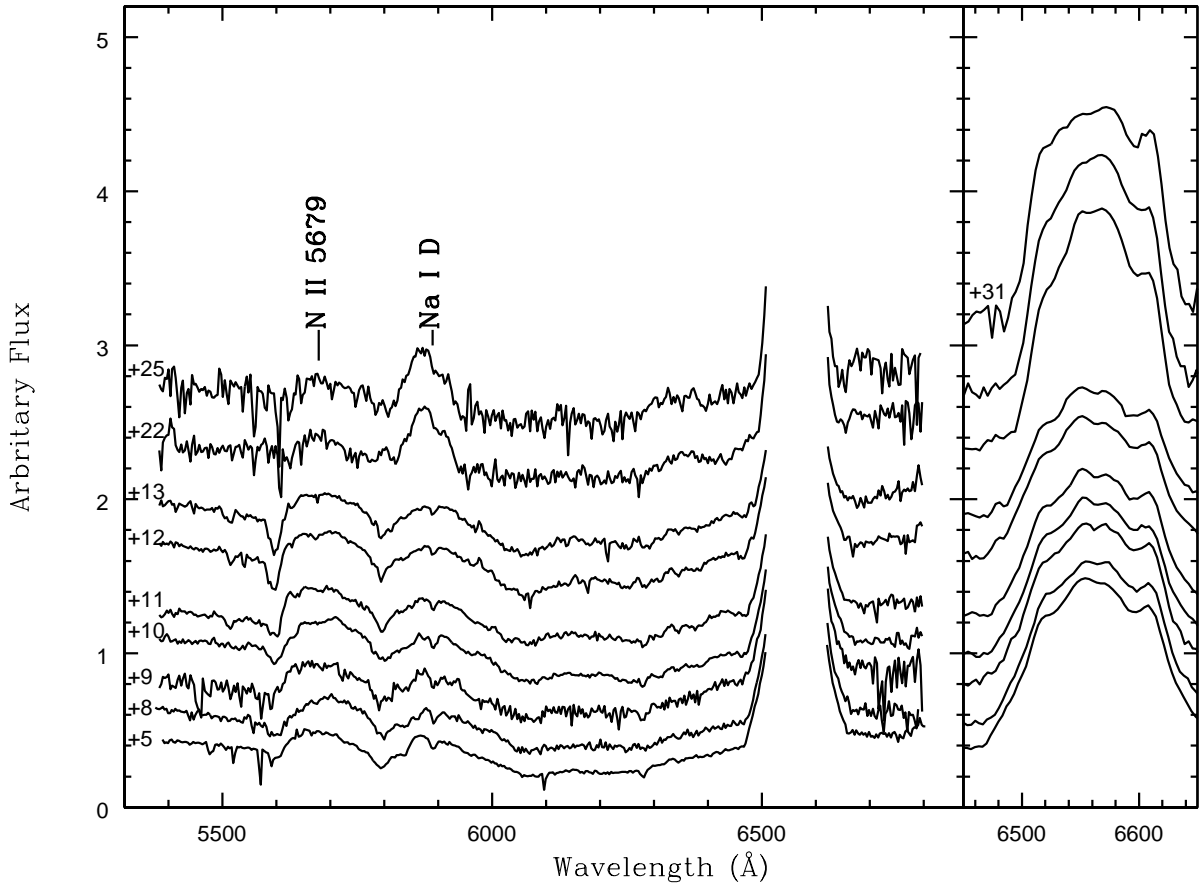}
  \caption{Meaburn spectrometric observations of V2491 Cyg during days 5$-$31 (2008 April 06 to 2008 May 12). The Meaburn spectra have been corrected for instrumental efficiency/atmospheric transmission using Be Star comparison spectra (see text for details). On the left-hand-side we have truncated the H$\alpha$ emission to show more clearly the lower intensity lines. The spectrum on day 31 after outburst was too noisy at this low level, so we only plot the H$\alpha$ line. On the right-hand-side the H$\alpha$ line profiles are shown. The time since outburst (in days) is shown to the left of each spectrum. Cosmic rays have also been removed. The higher resolution (non-Meaburn spectrograph) H$\alpha$ profiles in Table \ref{tb:spectra_obs} are shown in Fig. \ref{fig:spec_evo}.}
  \label{fig:combined}
\end{figure*}

The observations were carried out over several epochs from $t$ = 5 days (2008 April 16) to 31 days (2008 May 12) post-outburst. The log of spectrometric observations is shown in Table \ref{tb:spectra_obs}. The spectrograph provided three overlapping fixed grating positions, giving complete wavelength coverage from 3900$-$8000\AA. However, we only use a single grating position (5350$-$6888\AA) with dispersion of 3.00 \AA/pix and spectral resolving power $\sim$ 350. Data reduction was performed through a pipeline that initially performs bias, dark frame and flat field subtraction. With a comparison arc spectrum, a third order polynomial 2D wavelength/distortion map was created using the STARLINK Figaro ARC/IARC/ISCRUNCH routines and applied to the object image. Sky subtraction based on identification of non-object areas of the 2D spectrum was then carried out, if significant sky lines were visible, using the Figaro routine POLYSKY. A 1D spectrum was then automatically extracted using the PROFILE and OPTEXTRACT routines. Relative flux corrections were accomplished by using observations of a number of early type Be stars (showing few absorption lines) obtained for a different programme. Absorption lines were patched over and the star's temperature was chosen following their spectral types \citep{B81}. Using the STARLINK Figaro IRFLUX routine we corrected the spectra for instrumental efficiency/atmospheric absorption. In Fig. \ref{fig:combined} we show the computed flux-corrected spectra from $t$ = 5 to 31 days after outburst.

\begin{table}
  \caption{Log of optical spectral observations of V2491 Cyg.}
  \centering
  \begin{tabular}{lccc}
    \hline
    Date of & days after & Exposure & \multirow{2}{*}{Instrument}\\
    observation & outburst & time (sec) & \\
    \hline
    2008 April 13 & 2 & 900.0 & Varese-MMS \\
    2008 April 16 & 5 & 200.0 & Meaburn$-$LT \\
    2008 April 19 & 8 & 200.0 & Meaburn$-$LT \\
    2008 April 20 & 9 & 200.0 & Meaburn$-$LT \\
    2008 April 21 & 10 & 200.0 & Meaburn$-$LT \\
    2008 April 22 & 11 & 200.0 & Meaburn$-$LT \\
    2008 April 23 & 12 & 200.0 & Meaburn$-$LT \\
    2008 April 24 & 13 & 200.0 & Meaburn$-$LT \\
    2008 April 25 & 14 & 300.0 & SARG-TNG \\
    2008 May 03 & 22 & 600.0 & Meaburn$-$LT \\
    2008 May 06 & 25 & 600.0 & Meaburn$-$LT \\
    2008 May 12 & 31 & 600.0 & Meaburn$-$LT \\
    2008 May 14 & 33 & 1800.0 & Varese-MMS \\
    2008 July 27 & 108 & 900.0 & Asiago$-$AFOSC \\[1ex]
    \hline
  \end{tabular}
  \label{tb:spectra_obs}
\end{table}

We also obtained spectrometric observations at $t$ = 2, 25, 33, 108 and 477 days after outburst with: ($i$) the 0.6m telescope of the Schiaparelli observatory in Varese equipped with a multi mode spectrograph and various reflection gratings, ($ii$) the 3.5m TNG in La Palma and the high resolution spectrograph SARG ($\Delta \lambda / \lambda$ = 75000), and ($iii$) the spectrograph/imager AFOSC on the 1.82m telescope of the Padova Astronomical Observatory in Asiago with wavelength coverage of 6400$-$7050\AA. This is part of the observational monitoring of V2491 Cyg is described by \citet{MSD10} to which the reader is addressed for full details about data acquisition and reduction.

\section{Observational Results}\label{results}
We analysed the spectral information first to get an initial estimate of the component parameters before detailed modelling of V2491 Cyg began. The spectra show broad H$\alpha$, nitrogen (\nii\ 5679\AA) and sodium (\nai\ D) lines (Fig. \ref{fig:combined}). The spectral monitoring by \citet{MSD10} shows that at day +33 after outburst the auroral \niif\ 5755 line was still pretty weak, and thus the nebular \niif\ 6458, 6584 lines should not yet be perturbing the H$\alpha$ profile. Also evident is a narrowing of the H$\alpha$ line profile from day 22 after outburst, which is after the optical rebrightening. We cannot say why this rebrightening should happen but it is noteworthy that \citet{MSD10} also observed a decrease in the equivalent widths of the H$\beta$ line around this time. The \nii\ and \nai\ lines seen are similar to those seen in the early spectrum of V394 CrA \citep{WHP91}. However, these lines disappear from the spectrum of V394 CrA 20 days after its 1987 outburst. In the case of V2491 Cyg both are clearly present later after outburst. The \nii\ line, as in V394 CrA, is most likely due to continuum fluorescence from the post-outburst WD radiation.

\begin{figure}
  \centering
  \includegraphics[angle=270, width=75mm]{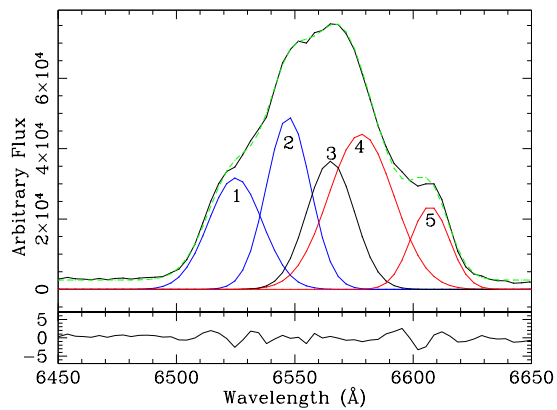}
  \caption{Gaussian components best fit to the observed H$\alpha$ line profile for V2491 Cyg on day 25 after outburst. The observed spectrum is shown in black and the combined Gaussian components in dashed green. The colour scheme for the individual Gaussian components are just to separate the different components. The component numbers, 1 through 5, are in increasing wavelength of the peak. Also shown are the residuals to the overall fit. The results for the derived radial velocities of each peak and FWHM of each component are presented in Tables \ref{tb:radial} and \ref{tb:fwhm}, respectively. (A colour version of this figure is available in the online journal.)}
  \label{fig:spectra_best_fit}
\end{figure}

The line of interest for our detailed analyses is H$\alpha$ because of its strength compared to those of other lines visible in the spectrum. Furthermore, later in outburst other emission lines are severely blended \citep{MSD10}. Using the low resolution Meaburn spectra we initially fit five Gaussian components to the spectra (see Fig. \ref{fig:spectra_best_fit}). This allows us to derive the individual components' radial velocities and their respective FWHM (see Tables \ref{tb:radial} and \ref{tb:fwhm}, respectively). There is evidence for velocity symmetry between the faster moving components, 1 and 5. Components 2, 3 and 4 are much harder to constrain, especially component 3 which becomes stronger after rebrightening. We find the displacement of the line profile from the rest wavelength of the H$\alpha$, using components 1 and 5, to be 92$\pm$36 \kms.
\begin{table*}
  \caption{Derived radial velocity for each of the 5 components of the H$\alpha$ line in V2491 Cyg (see Fig. \ref{fig:spectra_best_fit} and text for details).}
  \centering
  \begin{tabular}{lcccccc}
    \hline
    \multirow{2}{*}{Date} & days after & \multicolumn{5}{c}{Component Radial Velocity (\kms)} \\
    & outburst & 1 & 2 & 3 & 4 & 5 \\
    \hline
    2008 April 16 & 5 & -1875$\pm$32 & -530$\pm$34 & 117$\pm$3529 & 615$\pm$64 & 2028$\pm$68 \\
    2008 April 19 & 8 & -1860$\pm$17 & -510$\pm$22 & 112$\pm$961 & 671$\pm$61 & 2085$\pm$19 \\
    2008 April 20 & 9 & -1899$\pm$50 & -537$\pm$38 & 98$\pm$849 & 700$\pm$28 & 2088$\pm$3 \\
    2008 April 21 & 10 & -1828$\pm$65 & -482$\pm$48 & 99$\pm$2529 & 653$\pm$8 & 1994$\pm$53 \\
    2008 April 22 & 11 & -1909$\pm$76 & -531$\pm$55 & 92$\pm$1209 & 687$\pm$59 & 2043$\pm$24 \\
    2008 April 23 & 12 & -2025$\pm$163 & -642$\pm$93 & 120$\pm$1551 & 680$\pm$87 & 2072$\pm$120 \\
    2008 April 24 & 13 & -2054$\pm$67 & -720$\pm$72 & 90$\pm$1341 & 558$\pm$29 & 2009$\pm$65 \\
    2008 May 03 & 22 & -1634$\pm$47 & -610$\pm$30 & 184$\pm$29 & 673$\pm$30 & 2088$\pm$26 \\
    2008 May 06 & 25 & -1739$\pm$218 & -716$\pm$20 & 96$\pm$33 & 687$\pm$58 & 2015$\pm$44 \\
    2008 May 12 & 31 & -1865$\pm$92 & -822$\pm$111 & 334$\pm$91 & 368$\pm$368 & 2094$\pm$5 \\[1ex]
    \hline
  \end{tabular}
  \label{tb:radial}
\end{table*}

\begin{table*}
  \caption{Derived FWHM velocity for each of the 5 components of the H$\alpha$ line in V2491 Cyg (see Fig. \ref{fig:spectra_best_fit} and text for details).}
  \centering
  \begin{tabular}{lcccccc}
    \hline
    \multirow{2}{*}{Date} & days after & \multicolumn{5}{c}{Component FWHM Velocity (\kms)} \\
    & outburst & 1 & 2 & 3 & 4 & 5 \\
    \hline
    2008 April 16 & 5 & 1729$\pm$47 & 1357$\pm$24 & 1870$\pm$3487 & 1379$\pm$130 & 1166$\pm$91 \\
    2008 April 19 & 8 & 1664$\pm$48 & 1412$\pm$14 & 1807$\pm$1780 & 1429$\pm$109 & 1115$\pm$41 \\
    2008 April 20 & 9 & 1649$\pm$40 & 1399$\pm$101 & 1998$\pm$1222 & 1488$\pm$57 & 1050$\pm$27 \\
    2008 April 21 & 10 & 1658$\pm$101 & 1239$\pm$53 & 1995$\pm$2238 & 1261$\pm$137 & 1194$\pm$74 \\
    2008 April 22 & 11 & 1660$\pm$89 & 1283$\pm$59 & 2694$\pm$2978 & 1305$\pm$137 & 1084$\pm$53 \\
    2008 April 23 & 12 & 1639$\pm$180 & 1384$\pm$210 & 1617$\pm$776 & 1476$\pm$395 & 1030$\pm$167 \\
    2008 April 24 & 13 & 1421$\pm$110 & 1335$\pm$83 & 1765$\pm$2077 & 1364$\pm$169 & 1127$\pm$92 \\
    2008 May 03 & 22 & 1340$\pm$60 & 910$\pm$27 & 931$\pm$86 & 1502$\pm$105 & 806$\pm$40 \\
    2008 May 06 & 25 & 1232$\pm$255 & 996$\pm$236 & 1081$\pm$678 & 1477$\pm$382 & 875$\pm$63 \\
    2008 May 12 & 31 & 1189$\pm$103 & 1096$\pm$76 & 1725$\pm$298 & 1832$\pm$324 & 795$\pm$26 \\
    \hline
  \end{tabular}
  \label{tb:fwhm}
\end{table*}

\section{Modelling}\label{modelling}
Using {\sc shape} \citep[Version 3.56,][]{SL06,SKW10}\footnote{Available from http://bufadora.astrosen.unam.mx/shape/} we performed morpho-kinematical studies of V2491 Cyg. Classical nova ejecta have been modelled with various structures \citep[e.g.,][]{H72,S83,GO99}. Work by \citet{SOD95}, later updated by \citet{B02}, suggested a relationship between the speed class (in terms of the time for the wind flux to decline 3 magnitudes from peak $-$ $t_3$) and the major to minor axis ratio of the expanding nova shell, where the faster the nova the less the degree of shaping. This relationship works well with CNe (short orbital period systems) and possibly RNe of short orbital period. However, RS Oph, a system with orbital period of 455 days \citep{DK94,FJH00,BQM09}, at 155 days after outburst showed a deprojected major to minor axial ratio of 3.85 \citep{RBD09}, which considering its fast optical decline would not agree with the relationship originally found by \citet{SOD95}. Recent results for V2672 Oph also appear to disagree with this relationship \citep{MRB10}. However, we expect that the mechanism for formation of CN and RN shells, especially those RNe with long orbital periods, to be different. For example, systems such as RS Oph have remnants grossly affected by the interaction of ejecta with the pre-existing red-giant wind, which is unique to this sub-group of RNe.

Several mechanisms have been proposed to be involved in the shaping of CN ejecta. These can be identified as that associated with the common-envelope phase during outburst, the presence of a magnetised  WD, the spin of the WD and an asymmetric thermonuclear runaway \citep[see e.g.,][]{OB08}. The most widely accepted model involves the common-envelope phase where the ejecta from the WD envelope engulf the secondary star within a matter of minutes following outburst. The secondary transfers energy and angular momentum to the ejecta (\citealt{LSB90}; \citealt{LOB97}; \citealt{POB98}). WD rotation, when incorporated into calculations of the common-envelope phase, produces the observed prolate remnants \citep{POB98}.

\subsection{Model Assumptions}
It is desirable when dealing with modelling to reduce computing time and yet have a physically sound model. The modelling of permitted line profiles so early after outburst is complicated by the fact that they are most likely affected by an optically thick medium. The full treatment of radiative transfer requires full knowledge of several parameters of the system and further assumptions are required. For example, the temperature of the central system in outburst, which can be derived from multi-frequency observations, and the density of each component, which is much harder to derive since we must assume an ejected mass and a geometry of the system.
\begin{figure}
\resizebox{\hsize}{!}{\includegraphics[angle=270]{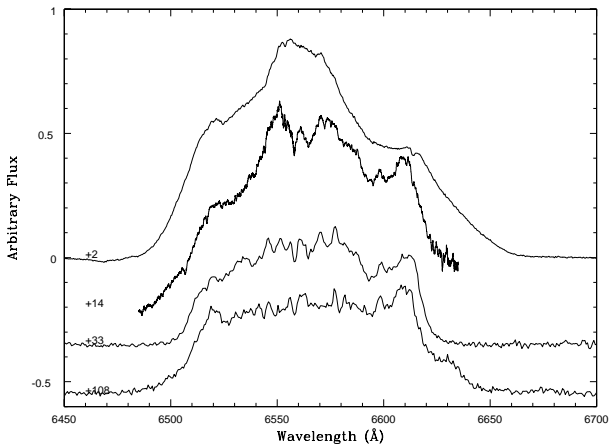}}
\caption{Spectral evolution of the high resolution H$\alpha$ line profile from several telescopes as shown in Table \ref{tb:spectra_obs} and in the text. The spectra has been offset and the numbers on the left are days after outburst.}
\label{fig:spec_evo}
\end{figure}

We considered the effects of radiative transfer on the line profile and found that the the lower the inclination the more the absorption affects the system but the overall line shape does not significantly change from the optically thin case. Furthermore, the absorption affects more the lower velocity systems. So when dealing with the results caution is necessary especially with the densities derived from the models because applying a density in the optically thin case will actually be only a lower limit to the true density of a system with greater opacity.

We also made the assumption that at early times the structures are still volume filled. \citet{VPD02} have shown the H$\alpha$ spectral evolution of V382 Vel from early outburst to the nebular stages (5 to 498 days after outburst); at early stages the line profile showed an asymmetric saddle-shape with the blue component more prominent than the red. The blue component diminishes in strength and the whole line profile develops to become a flat-topped shape. They associate this with the termination of the post-outburst wind phase and the complete ejection of the envelope. Our higher resolution spectra, Fig. \ref{fig:spec_evo}, also show a similar development, and a similar interpretation can be made here.

\subsection{Model Parameters}
\begin{figure}
\resizebox{\hsize}{!}{\includegraphics{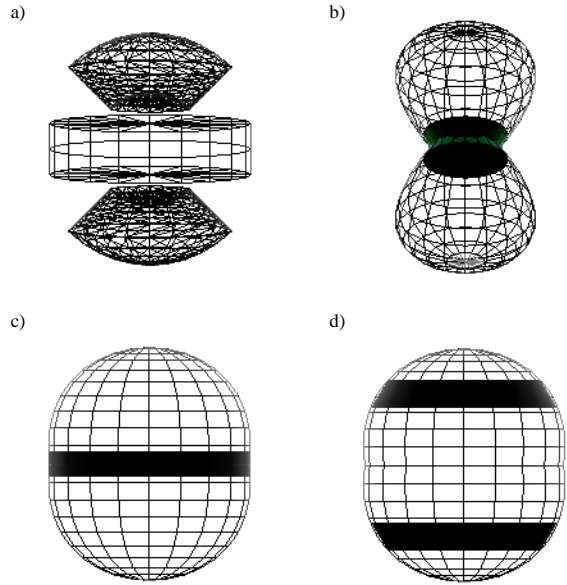}}
\caption{Models as visualised in  {\sc shape}. a) polar blobs with an equatorial ring (Model A), b) a dumbbell structure with an hour-glass overdensity (Model B), c) prolate shell with an equatorial ring (Model C) and d) prolate shell with tropical rings (Model D).}
\label{fig:models}
\end{figure}

Given V2491 Cyg's fast decline from maximum we assume an axial ratio of 1.15 for our chosen structures following the relationships in \citet{SOD95}. We also consider the possibility of a structure similar to RS Oph, first coined ``peanut-shaped'' \citep{BHO07} and later modelled as a dumbbell structure with an hour-glass-shaped central overdensity \citep{RBD09}. This model is included in our studies because, as mentioned before, this system has been suggested as a RN and a similar structure has been proposed for the RN U Sco \citep{DO10}. The candidate structures we chose are (i) polar blobs with an equatorial ring \citep[Model A, e.g., HR Del; ][]{H72,S83}; (ii) a dumbbell with an inner hour-glass overdensity with an axial ratio the same as derived above for RS Oph (Model B); (iii) prolate shell with an equatorial ring (Model C); and (iv) a prolate shell with tropical rings \citep[Model D, e.g., V705 Cas; ][]{GO99}. As noted above, Model B differs from that generated by the common-envelope phase described above because the observed structure could arise due to interaction with a pre-existing red-giant wind and/or collimation by the accretion disc (e.g., \citealt{BHO07}; \citealt{SRM08}; \citealt{RBD09}). Our models, as visualised in {\sc shape}, are shown in Fig. \ref{fig:models}.

The lack of a resolved image of the remnant means any model is not as well constrained as when the structure is spatially resolved. However, for the models assumed above we constrain a system inclination (where an inclination $i$ = 90 degrees corresponds to the orbital plane being edge-on, and $i$ = 0 degrees being face-on) and an expansion assumed to take place in a Hubble flow.

We explore the parameter space from 0$-$90 degrees and maximum expansion velocity ($V_{\textrm{exp}}$) ranging from 100$-$8000 \kms\ (in steps of 1 degree and 100 \kms, respectively). We exclude $V_{\textrm{exp}}$ $>$ 8000 \kms\ as this is much higher than those normally associated with novae and indeed is at the lower end of velocities found in Type Ia Supernovae \citep[e.g.,][]{L00}. We note here however, that in the helium nova V445 Pup high speed knots were observed moving at 8450$\pm$570 \kms\ \citep{WSK09}. The models were run several times to produce a well sampled model spectrum. The model spectra are compared to the observed spectra and flux matched via $\mathcal{X}^2$ minimisation, using techniques from \citet{PTV92}. Furthermore, an estimate of the density ratio of each component was derived from their total fluxes using the task {\sc specfit} in IRAF\footnote{IRAF is distributed by the National Optical Astronomical Observatories, which is operated by the Associated Universities for Research in Astronomy, Inc., under contract to the National Science Foundation.}. These were used as inputs for the relative structure densities.

\section{Modelling Results}\label{results2}
\subsection{First epoch observations at $t$ = 25 days after outburst}
\begin{figure*}
\centering
\includegraphics[width=15cm]{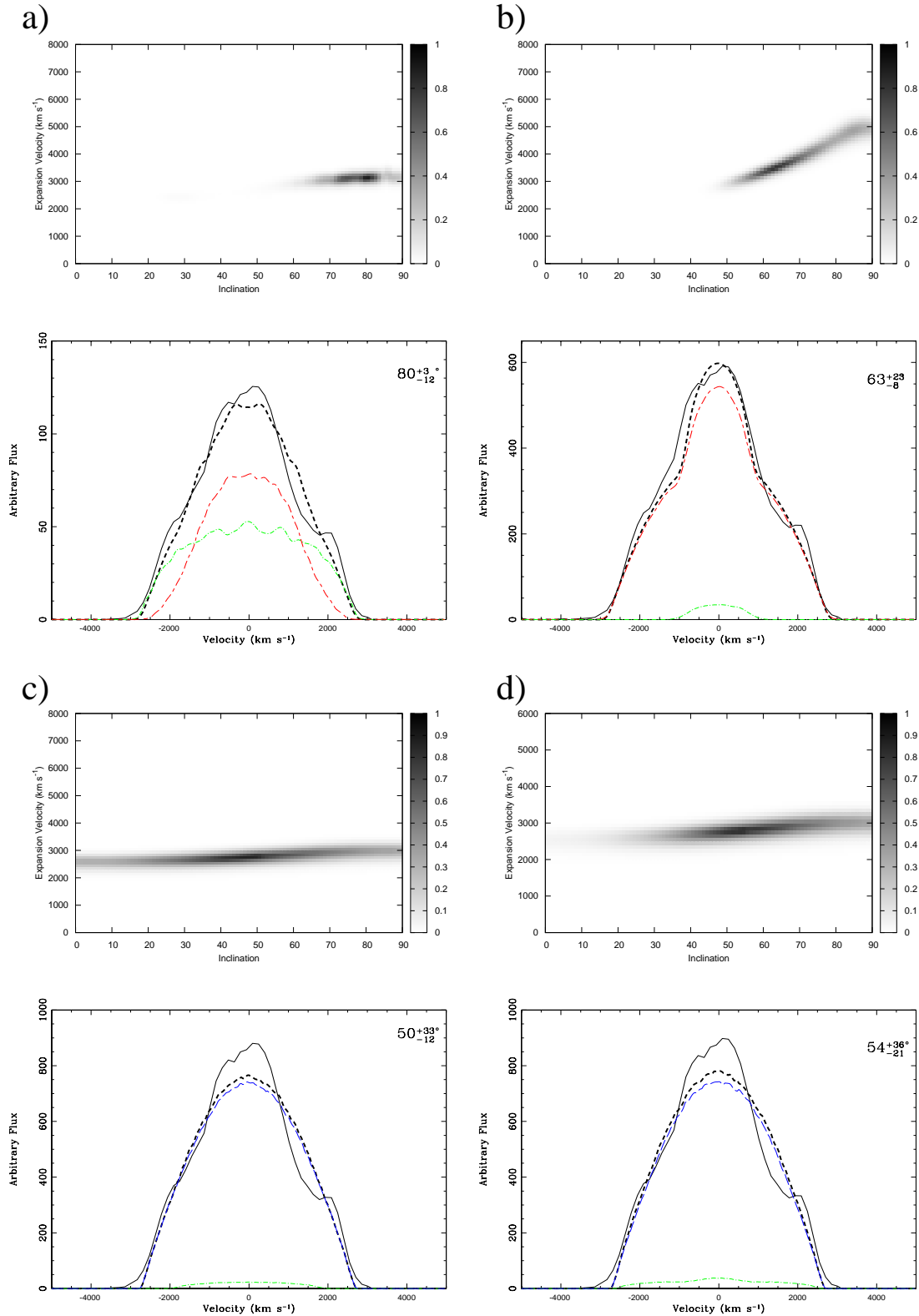}
\caption{Results of the best fit model spectrum to the observed spectrum on day 25 after outburst. The images are for a) polar blobs with an equatorial ring (Model A), b) dumbbell structure with an hour-glass overdensity (Model B), c) prolate shell with an equatorial ring (Model C) and d) prolate shell with tropical rings (Model D). Displays, top $-$ the most likely result for the structure, where the colour gradient represents the probability that the observed $\mathcal{X}^2$ value is correct. Bottom $-$ the observed (solid black) and modelled spectrum (short-dash black). Also shown are the individual component contributions; rings (dot-dash green), prolate (long-dash blue) and polar blobs and dumbbell (short-dash-long-dash red). (A colour version of this figure is available in the online journal.)}
\label{fig:model_result}
\end{figure*}

Shown in Fig. \ref{fig:model_result} are the results of our model fits, and their respective 1$\sigma$ errors, for all the model runs. Panel a) shows the results for Model A. These suggest that the best fit for this model is an inclination of 80$^{+3}_{-12}$ degrees with $V_{\textrm{exp}}$ for the polar blobs of $\sim$ 3100$^{+200}_{-100}$ \kms\ and for the equatorial ring of $\sim$ 2700$^{+200}_{-100}$ \kms. The individual component fluxes are about the same. Panel b) shows the results for Model B. These suggest that the best fit to the model is for a system with inclination 63$^{+23}_{-8}$ degrees and $V_{\textrm{exp}}$ of 3500$\pm$$^{+1500}_{-300}$ \kms. The flux is dominated by the dumbbell component. Panel c) shows the results for Model C. These suggest that the best fit to the model is for a system with inclination 50$^{+33}_{-12}$ degrees and $V_{\textrm{exp}}$ of 2800$^{+300}_{-200}$ \kms. The flux is dominated by the prolate structure. Finally, panel d) shows the results for Model D. These suggest that the best fit to the model is for a system with inclination 54$^{+36}_{-21}$ degrees and $V_{\textrm{exp}}$ of 2800$^{+300}_{-200}$ \kms. The flux is again dominated by the prolate structure. 

As previously mentioned these results are not as well constrained as they would be if we could have the spectra combined with imaging data. We also performed modelling for a simple sphere. The results suggest that the best fit maximum expansion velocity $V_{\textrm{exp}}$ $\sim$ 2700 \kms. The shape of the synthetic spectra was that of an inverted U (similar to the prolate components in Models C and D of Fig. \ref{fig:model_result}).

To determine the best fitting morphology we examined the derived reduced $\mathcal{X}^2$ probabilities and found that Models A and B showed comparable probabilities. The other structures gave negligible probabilities.

\subsection{Second epoch observations at $t$ = 108 days after outburst}
Another test we perform on our models to see which best reproduces the data, considering only Models A and B, is to evolve the models to a later epoch, keeping the parameters for inclination and $V_{\textrm{exp}}$ as derived from the first epoch to the second epoch. In the case of Model A, it was allowed to evolve linearly. However, \citet{RBD09} suggested that their central component for RS Oph showed deceleration and we therefore kept this central component the same size from one epoch to the next in Model B.

What is evident is that if the models are allowed to evolve linearly, with the derived parameters the same, then the line profile should look identical from one epoch to the next in the absence of other physical changes. For the line profile to change, but with the velocity field and inclination constant, there must be a transition involving other factors. Just changing the densities does not suffice without the addition of new components. This transition may therefore be associated to the termination of the post-outburst wind phase and the complete ejection of the envelope, which is proposed to happen within the first couple of weeks \citep[as in V382 Vel; ][]{VPD02}. Therefore, the later epoch modelling does not have a filled distribution but now has a depth of material (shell) determined by the first epoch model. We therefore apply a volume distribution within this shell to both models and compare with the spectrum at day 108 after outburst (Fig. \ref{fig:spec_evo}). Furthermore, \citet{MSD10} from photo-ionisation models showed the nova to be in the nebular stage at day 108 after outburst and the \niif\ 6548 and 6584\AA\ emission lines may contribute significantly to the H$\alpha$ profile. Certainly \oiii\ was about ten times stronger than H$\alpha$ at this time \citep[see Table 4 in][]{MSD10}.

\subsubsection{Model B $-$ Dumbbell with an hour-glass overdensity}
We do not show the Model B synthetic spectrum as it did not reproduce the observed second epoch spectrum at all well. The synthetic spectrum was that of an upside down ``T'' shape. We therefore, believe that Model A is the correct model for V2491 Cyg.
\begin{figure*}
  \centering
  \resizebox{\hsize}{!}{\includegraphics{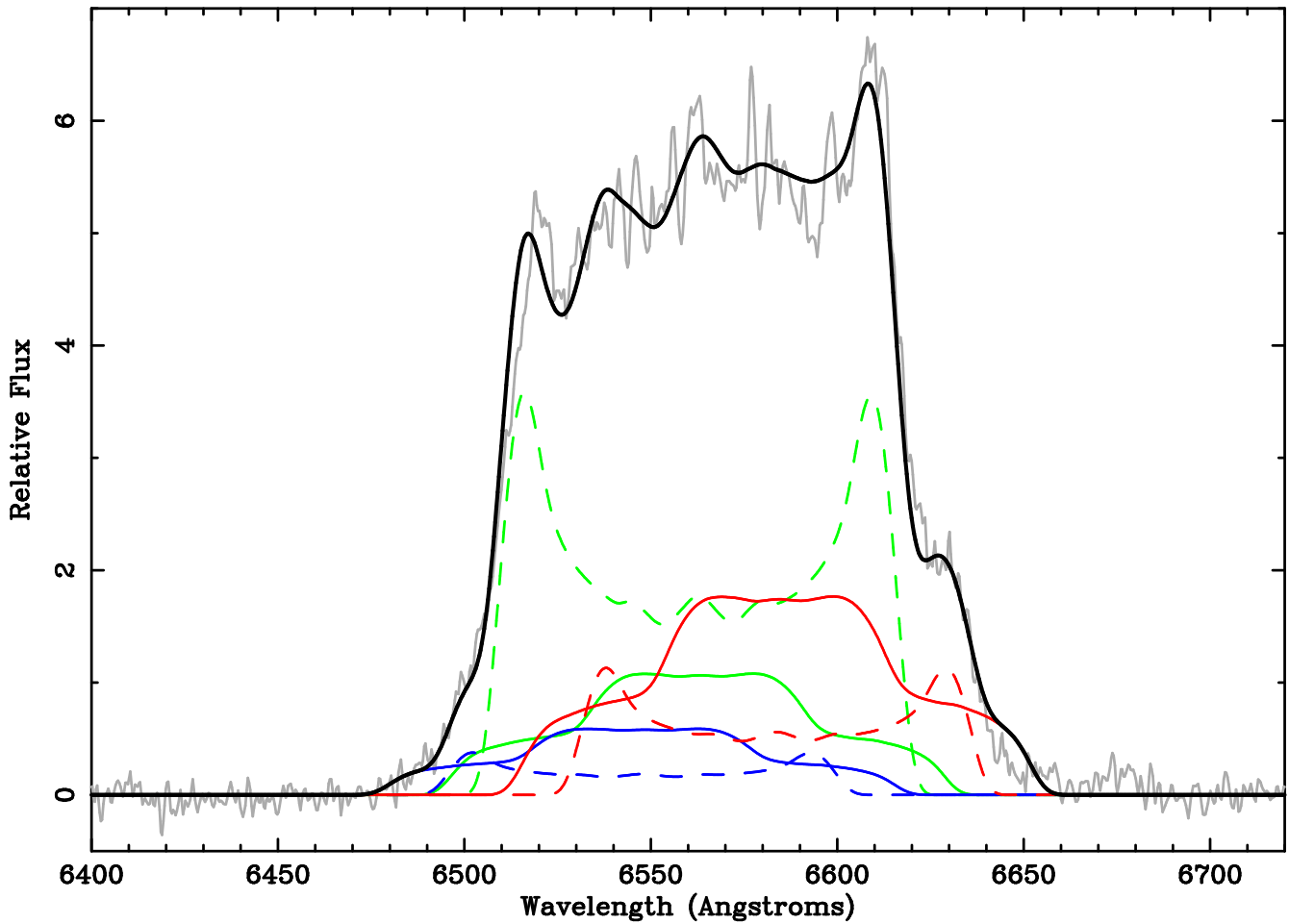}}
  \caption{Fit to the observed line profile (grey) at day 108 after outburst. The model spectrum (black) is the sum of the individual components, H$\alpha$ (green), \niif\ 6584\AA\ (red) and \niif\ 6548\AA\ (blue). The ring (dashed line) and polar blobs (solid line). The ratio of the \niif\ lines was kept at 1:3 and the relative flux of H$\alpha$ to the \niif\ lines was kept as a free parameter. (A colour version of this figure is available in the online journal.)}
  \label{fig:epoch2}
\end{figure*}

\subsubsection{Model A $-$ Polar blobs and equatorial ring}
In light of the presence of \niif\ we should account for this when modelling the spectra on day 108. We should expect \niif\ to appear as the expanding remant becomes less dense overall and the nebular spectrum emerges. In the first instance, to fit the observed spectrum the velocities were kept constant from the first epoch for the H$\alpha$ and \niif\ lines and only the relative flux was allowed to change. We assume here that the model derived for H$\alpha$ is the same for \niif. This did not reproduce the observed spectrum without introducing significant additional absorption features.

We therefore allowed the velocity of the different components to change keeping the flux ratio of the \niif\ lines always at 1:3 and the relative flux of the H$\alpha$ to the \niif\ lines as a free parameter. The result of this modelling is shown in Fig. \ref{fig:epoch2}. This provides an excellent fit to the spectrum at day 108. The derived velocities for the polar blobs were $\sim$ 3500 \kms\ (H$\alpha$) and $\sim$ 3600 \kms\ (\niif) and for the equatorial ring $\sim$ 2700 \kms\ (H$\alpha$) and $\sim$ 2600 \kms\ (\niif). These results are within 2$\sigma$ of those derived for the first epoch.

It should be noted that there is a change in the contributions to the flux of the different components of H$\alpha$. Where in the first epoch the two components were of comparable flux, in the second epoch the ring is the major contributor to the flux with relative flux contribution of the polar blobs to the ring of 1:2.5 for H$\alpha$. This change in relative flux can be easily understood as arising from the fact that material moving at higher velocities will fade faster in H$\alpha$ emission. Therefore, the change shown in Fig. \ref{fig:epoch2}, where the equatorial ring becomes the dominant component stems from the fact that it presents an expansion velocity lower than that of the polar blobs.

In the case of the \niif\ line the polar blobs to ring relative flux contribution is 2.2:1. Their expansion velocities are $\sim$ 3600 \kms\ and $\sim$ 2600 \kms, respectively. We should also note that \citet[][their Figure 10]{MSD10} show a line profile at day 477 which is very different from the day 108 line profile. It does not have a flat top profile and has significant asymmetry. Our model can also account for this if the densities of the different components continue to evolve. As the nebula becomes less dense we expect that the relative contribution from \niif\ should increase.

There is one more test we can perform to further constrain our model. The derived inclination of 80 degrees is very high and we therefore expect eclipses whose existence we are now investigating. Such eclipses should be easily observed in this system. For example, the CN DQ Her (orbital period $\sim$ 4 h) shows deep $\sim$ 0.9 mag eclipses in UBVR \citep{ZRS95} as does the RN U Sco with eclipses of depth $\sim$ 1.5 mag in B \citep{SR95}. Of course, if V2491 Cyg turns out to be eclipsing this will also enable us to determine more precisely important parameters of the central binary.

\section{Discussion and Conclusions}\label{discussion}
Of the morphologies discussed above early in outburst, Models A and B, are the best fitting and give results that are very similar to each other early in outburst and this is reflected on the fact that the inclinations of the two morphologies overlap, within the errors. Furthermore, the main real difference is in the size of the central region. Model A is representative of a CN system, while the Model B has a morphology that has been observed in the RN RS Oph.

To model the second epoch observations the model was allowed to evolve into a system that was not volume filling but a shell. To replicate the observed spectrum with Model A at this time we needed to introduce significant emission from the \niif\ lines and the components densities required to be changed with the equatorial ring becomimg the dominant contributor to the flux. The change in derived densities between the components should also be expected since the different components are moving at different speeds. Emission from the polar blobs moving at higher speed will be observed to fade faster than the equatorial ring. Model B at the second epoch did not reproduce the line profile. Therefore, from the second epoch observations we suggest that Model A is the most likely with an inclination of 80$^{+3}_{-12}$ degrees and a maximum expansion velocity of 3100$^{+200}_{-100}$ \kms, for the polar blobs. This velocity is in line with those found in high-ionisation absorption lines in X-ray spectra by \citet{N10}.

The high inclination found for Model A is however difficult to reconcile with the outburst amplitude vs. decline rate relationship for classical novae \citep{W87}. The observed amplitude of 10 mags is much lower than that predicted for the derived inclination of 80 degrees (15.5 mags) while a pole-on system would still have a predicted amplitude of 12.5 mags. However, we note that RNe present a similar outburst amplitude at a similar inclination \citep[e.g.][]{MZT99}. This is in line with the suggested RN nature of this object from other authors \citep[e.g.,][]{POE10}. Further exploration of the nature of the progenitor will be presented in \citet{DRB10}.

\section*{Acknowledgments}
The authors are grateful to W. Steffen and N. Koning for valuable discussions on the use of {\sc shape} and adding special features to the code, N. Clay and C. Mottram for useful discussions on programming, Alessandro Siviero, Paolo Valisa and Lina Tomasella for assisting in collecting the spectra in Asiago, Witold Maciejewski for discussion on radiative transfer and Chris Collins for discussions on statistics. VARMR is funded by an STFC studentship. The Liverpool Telescope is operated on the island of La Palma by Liverpool John Moores University in the Spanish Observatorio del Roque de los Muchachos of the Instituto de Astrofisica de Canarias with financial support from the STFC. We thank an anonymous referee for valuable comments on the original manuscript.


\label{lastpage}

\end{document}